\documentclass[
reprint,
amsmath,
pra,
]{revtex4-1}
\usepackage{tocvsec2}
\usepackage{float}
\usepackage{graphicx}
\usepackage{dcolumn}
\usepackage{hyperref}
\hypersetup{
colorlinks=true,
linkcolor=blue,
filecolor=magenta,
urlcolor=cyan,
}
\begin{document}

\title{Doppler cooling thermometry of a multi-level ion in the presence of micromotion}

\author{Tomas Sikorsky}
\email{tomas.sikorsky@weizmann.ac.il}
\author{Ziv Meir}
\author{Nitzan Akerman}
\author{Ruti Ben-shlomi}
\author{Roee Ozeri}
\affiliation{Department of Physics of Complex Systems, Weizmann Institute of Science.}

\date{\today}
\begin{abstract}
We study the time-dependent fluorescence of an initially hot, multi-level, single atomic ion trapped in a radio-frequency Paul trap during Doppler cooling. We have developed an analytical model that describes the fluorescence dynamics during Doppler cooling which is used to extract the initial energy of the ion. While previous models of Doppler cooling thermometry were limited to atoms with a two-level energy structure and neglected the effect of the trap oscillating electric fields, our model applies to atoms with multi-level energy structure and takes into account the influence of micromotion on the cooling dynamics. This thermometry applies to any initial energy distribution. We experimentally test our model with an ion prepared in a coherent, thermal and Tsallis energy distributions. 
\end{abstract}

\maketitle
\tableofcontents
\section{Introdution}
Doppler cooling of atoms was proposed by Hansch and Schawlow and independently by Wineland and Dehmelt in 1975 \cite{Hansch1975,Wineland1975}. Here, laser light is tuned such that, due to the Doppler effect, an atom moving against the laser propagation direction has a higher probability of scattering photons than an atom moving along the laser direction. Photon scattering thus cools the atom until it reaches the Doppler temperature limit, typically in the mK range. The ability to cool trapped ions to mK temperatures is a requirement for many applications, such as quantum information \cite{Leibfried2003}, optical clocks \cite{Margolis2009} and quantum simulations \cite{Blatt2012}. Doppler cooling is also a pre-requisite for sideband cooling in which the ion is cooled to its motional ground-state. 

Recently, the dynamics of recorded fluorescence during Doppler cooling of an initially hot ion was proposed and demonstrated as a tool for the evaluation of the ion energy \cite{Epstein2007, Wesenberg2007}. This method is widely used for thermometry of ions \cite{HAFFNER2008,Allcock2010,Blakestad2009,Steiner2013,Zipkes2010,Huber2008,Walther2012,Zipkes2010a,Abah2012,Daniilidis2011}. The Doppler cooling thermometry used in these experiments assumes a simple two-level cooling transition. However, in many of these experiments, the cooling transition involved multiple levels in a $\Lambda$ transition structure. The use of a two-level model produces inaccurate results.

While for ions with a simple two-level energy structure, both theory of Doppler cooling \cite{Cirac1992} and its use for thermometry are well established \cite{Wesenberg2007}, an accurate analytic model for cooling dynamics involving multiple levels in a $\Lambda$ transition structure does not exist. Such complex energy level structure has benefits in reaching sub Doppler-limit temperatures or even ground-state cooling as in EIT cooling schemes \cite{Lechner2016}. Recently, a new thermometry method which uses a dark-resonance in multi-level ions was proposed \cite{Roßnagel2015}. However, such dark resonance thermometry is only applicable up to tens of mK.

A further complication to the modeling of Doppler cooling and thermometry in Paul traps is the distortion of the absorption spectrum by micromotion. Micromotion is a fast, ion motion driven by the time-dependent trapping radio-frequency (rf) field. Excess micromotion can be eliminated by positioning the ion in the position where rf fields nulls \cite{Berkeland1998}. Inherent micromotion is always present due to the finite amplitude of ion's motion. Previous works on Doppler cooling theory treated the effect of micromotion by including the sidebands in the absorption spectra in the low-saturation limit \cite{Wesenberg2007}, or by restricting to the well-resolved micromotion sidebands regime \cite{Cirac1994}.

In this paper, we present an analytical Doppler cooling model applicable to multi-level ions, trapped in a radio-frequency Paul trap. We consider the effect of both excess and inherent micromotion outside the low-saturation limit. We also do not restrict our treatment to resolved micromotion sidebands. Our model predicts the time dependence of both energy and fluorescence during Doppler cooling. Using this model we obtain the initial energy distribution of the ion from an experimental time-resolved Doppler cooling fluorescence signal. We test our model using a single $\mathrm{Sr}^{+}$ ion in a linear Paul trap. The energy levels of $\mathrm{Sr}^{+}$ form a typical $\Lambda$ system with an eight-level manifold due to the Zeeman splitting under a constant magnetic field (\autoref{FIG. 1.}). Although our method applies to any level structure, without loss of generality, throughout this paper we will deal specifically with the level structure of $\mathrm{Sr}^{+}$, shown in \autoref{FIG. 1.}.

\begin{figure}
\includegraphics[width=\linewidth]{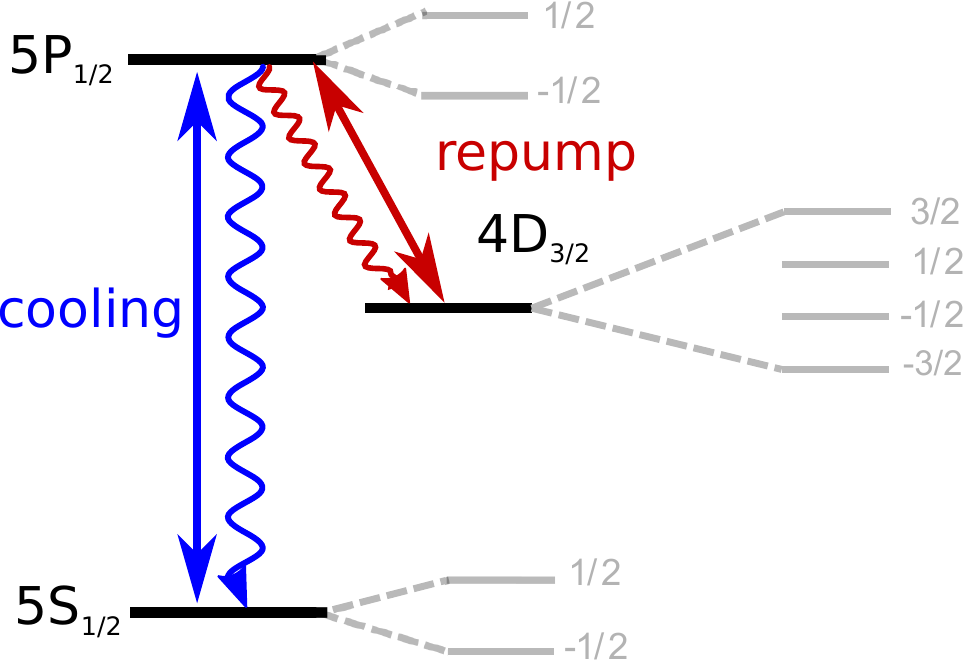}
\caption{\label{FIG. 1.} Relevant Energy levels of $^{88}\textrm{Sr}^+$. This $\Lambda$ level scheme is typical to several alkali-earth metal ions. The $4D_{3/2}$ is meta-stable with a lifetime of 390 msec. Both $S_{1/2}\rightarrow P_{1/2}$ and $D_{3/2}\rightarrow P_{1/2}$ are dipole allowed transitions. The branching ratio of the excited state decay into the $4D_{3/2}$ level is 1:17 and is given by the ratio of the natural linewidths of the associated transitions \cite{Zhang2016}.}
\end{figure}

\section{Doppler cooling thermometry}
Doppler cooling thermometry is based on the fact that the ions' kinetic energy influences the fluorescence rate due to the Doppler effect. To determine the initial energy of an ion, we monitor its fluorescence during Doppler cooling. The ion interacts with a red detuned, linearly polarized, laser light. As the ion scatters photons, its energy decreases towards the Doppler cooling limit and its fluorescence rate increases. Once the ion cools to an energy where Doppler shifts are too small to induce an appreciable change to the scattering rate, Doppler cooling thermometry loses its sensitivity. In our case, the transition on which cooling occurs is $\gamma = 2\pi \times 20.4\ MHz$ and the Doppler cooling signal loses sensitivity below $\approx 10\ mK$. The effect of the energy quantization of the ions motional levels is only important close to the trap ground-state and can be safely ignored in our analysis.

\subsection{The eight-level system}
For a two-level system, the steady state scattering rate can be expressed analytically \cite{Wesenberg2007}, however, for more than two-levels, there is no closed analytic formula. We numerically calculate the steady-state scattering rate of $^{88}\textrm{Sr}^{+}$ by solving the coupled Bloch equations for all involved levels with given laser couplings and decay channels (see SM II.). Here the levels we consider are the $5S_{1/2}$ ground-state and the $4D_{3/2}$ meta-stable state, with a lifetime of $390\ \mathrm{msec}$. These levels are connected via allowed dipole transitions to the $5P_{1/2}$ excited state at transition wavelengths of $422\ \mathrm{nm}$ and $1092\ \mathrm{nm}$ respectively. The lifetime of the $5P_{1/2}$ level is $8\ \mathrm{nsec}$. Including all Zeeman states of these levels we get 64 coupled Bloch equations (see \autoref{FIG. 1.}). We calculate the scattering rate, $\gamma=\Gamma\rho\left ( P_{1/2} \right )$, from the steady state population of the excited $5P_{1/2}$ states. To experimentally calibrate our laser couplings, we measure the scattering rate as a function of the $422\ \mathrm{nm}$ laser detuning (\autoref{FIG. 2.}). From a fit to the spectrum we find the Rabi frequencies and exact detuning of both lasers. As seen, the experimental data agrees well with the numerical solution.

With all laser parameters determined, we use the calculated spectrum to estimate the ion scattering rate at any given velocity. Here the Doppler shifts from both transitions are related through $\delta_{D}=\delta_{422}=\frac{1092}{422}\delta_{1092}$, where, $\delta_{422}=\mathbf{k_{422}\cdot v}$ and $\delta_{1092}=\mathbf{k_{1092}\cdot v}$ and we assume that both lasers are co-linear. The green solid line in \autoref{FIG. 22.} shows the calculated Doppler shifted spectrum of the ion. 

\begin{figure}
\includegraphics[width=\linewidth]{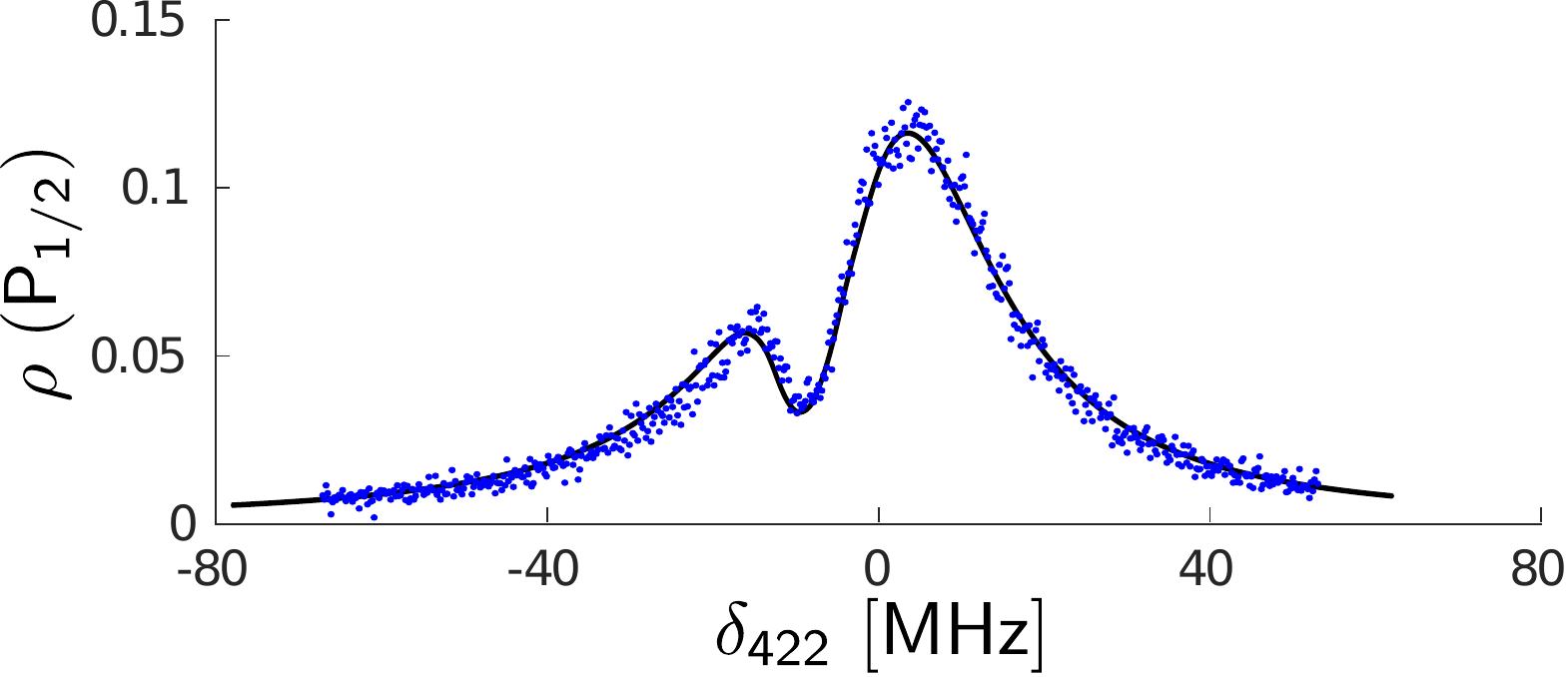}
\caption{\label{FIG. 2.}Spectroscopic scan of the $S_{1/2}\rightarrow P_{1/2}$ transition. We scan the 422 nm laser detuning, $\delta_{422}$. Blue points are the number of photons detected on a Photon-counter which is proportional to the excited state population $\rho(P_{1/2})$. The black line is a fit to the steady-state solution of the 8-level optical Bloch equations, which yields the following laser parameters: $\Omega_{422}/2\pi=9.3\pm 0.07$ MHz, $\Omega_{1092}/2\pi=7.0\pm 0.32$ MHz, $\delta_{1092}=-9.0\pm 0.14$ MHz. Additional parameters used to obtain the steady-state solution were measured independently (see SM I.).}
\end{figure}
\subsection{Weak binding limit}
Even with the fluorescence spectrum in hand, solving the energy-time-dependent problem of the scattering rate of a trapped ion in a laser field is a computationally intensive task. One has to solve the optical Bloch equations for an eight-level system which are also coupled with motion due to Doppler shifts to get the cooling rate. To develop a computationally efficient model which also gives a physical insight into the dynamics of Doppler cooling, several approximations have to be made. The first approximation is that of weak binding, which assumes that level populations, and therefore the fluorescence rate, reach steady-state faster than the ion is moving in the trap. This approximation was also used in the case of a two-level ion \cite{Wesenberg2007,Epstein2007}.

In the case of two-level atoms \cite{Wesenberg2007,Epstein2007} where the two levels are connected with a strong dipole transition (natural linewidth of several 10's MHz), the time scale over which populations reach steady state is given by the lifetime of the excited state and is independent of the laser intensity or detuning \cite{Noh2010}. For a multi-level atom, Doppler cooling involves additional re-pump transitions; the 1092 nm transition in our case. Typically, these transitions have natural linewidth about an order of magnitude smaller. The time scale over which populations reach steady state will depend on the intensity of the re-pump laser and will typically be an order of magnitude slower.

\subsection{Harmonic trap}
In this section, we concentrate on the effect of multi-level structure on Doppler cooling and ignore the effect of both inherent and excess micromotion. In linear Paul traps, the electric field in the axial direction is static. The simplified model of this section is appropriate for the determination of the ions' energy along the axial mode. The effects of micromotion and the extension of our model to motion in the radial directions are discussed in the following sections.

Similarly to \cite{Wesenberg2007}, we consider an event in which the atom scatters a single photon. From conservation of energy and momentum, we can show \cite{Wineland1979} that the energy of the ion in each mode $(i=x,y,z)$ per single scattering event will change by:
\begin{equation}
\Delta E_{i}=\hbar \left (k_{i}^{inc}-k_{i}^{scatt} \right )v_{i}+\frac{1}{3}\frac{\left (\hbar \left |k \right | \right )^2}{2m}\left (\hat{k}^{inc}-\hat{k}^{scatt} \right )^2. \label{delta1}
\end{equation}
Here, $v_i$ is the initial velocity of the ion, $k_i^{inc}$ is the incident photon $k$-vector and $k_i^{scatt}$ is the scattered photon $k$-vector projected on the $i$-th axis. In the non-relativistic limit, the magnitude of the k-vector changes by order of $\sim v/c$ which is small and can be neglected.
Since the re-scattered photon is isotropically emitted \cite{Letokhov1976}, we can integrate $k_i^{scat}$ over all scattering angles and \autoref{delta1} simplifies to,
\begin{equation}
\Delta E_{i}=\hbar k_{i}^{inc}v_{i}+\frac{2}{3}\frac{\left (\hbar \left | k \right | \right )^2}{2m}. \label{delta2}
\end{equation}

The cooling rate is now given by $\frac{\mathrm{d} E_{i}}{\mathrm{d} t}=\gamma \left ( \mathbf{par},\delta_D \right )\Delta E_{i}$, where $\gamma$ is the instantaneous scattering rate which we calculated by solving the eight-level optical Bloch equations. The scattering rate depends naturally on the Doppler shift, $\delta_{D}$, and also on $\mathbf{par}$ which is a vector of experimental parameters such as laser frequencies and intensities, magnetic field and more (see SM I.). These parameters are energy independent and therefore do not change during the cooling process.

In the weak binding regime, the internal states of the ion reach steady-state much faster than the typical time scale determined by the trap. The steady state population and the instantaneous scattering rate depend on the laser parameters and the instantaneous velocity of the ion. The velocity probability distribution along a single dimension for a harmonic oscillator which also determines the Doppler shift distribution is given by,
\begin{equation}
P_{1\textrm{D}}^{i}\left ( \delta _{M}^{i}, \delta ^{i} \right )=\tfrac{1}{\pi\sqrt{(\delta _{M}^{i})^2-(\delta^i)^2}}. \label{PSP}
\end{equation}
Here, $\delta_{M}^{i}$ is the maximal Doppler shift for a given ion energy, $E_i$, in a given trap direction $(i=x,y,z)$, $\delta _{M}^{i}=\frac{k_{i}}{2\pi}\sqrt{\frac{2E_{i}}{m}}$ and $k_i$ is the projection of the laser k-vector on a given trap axis.

\begin{figure}
\vspace{2mm}
\includegraphics[width=\linewidth]{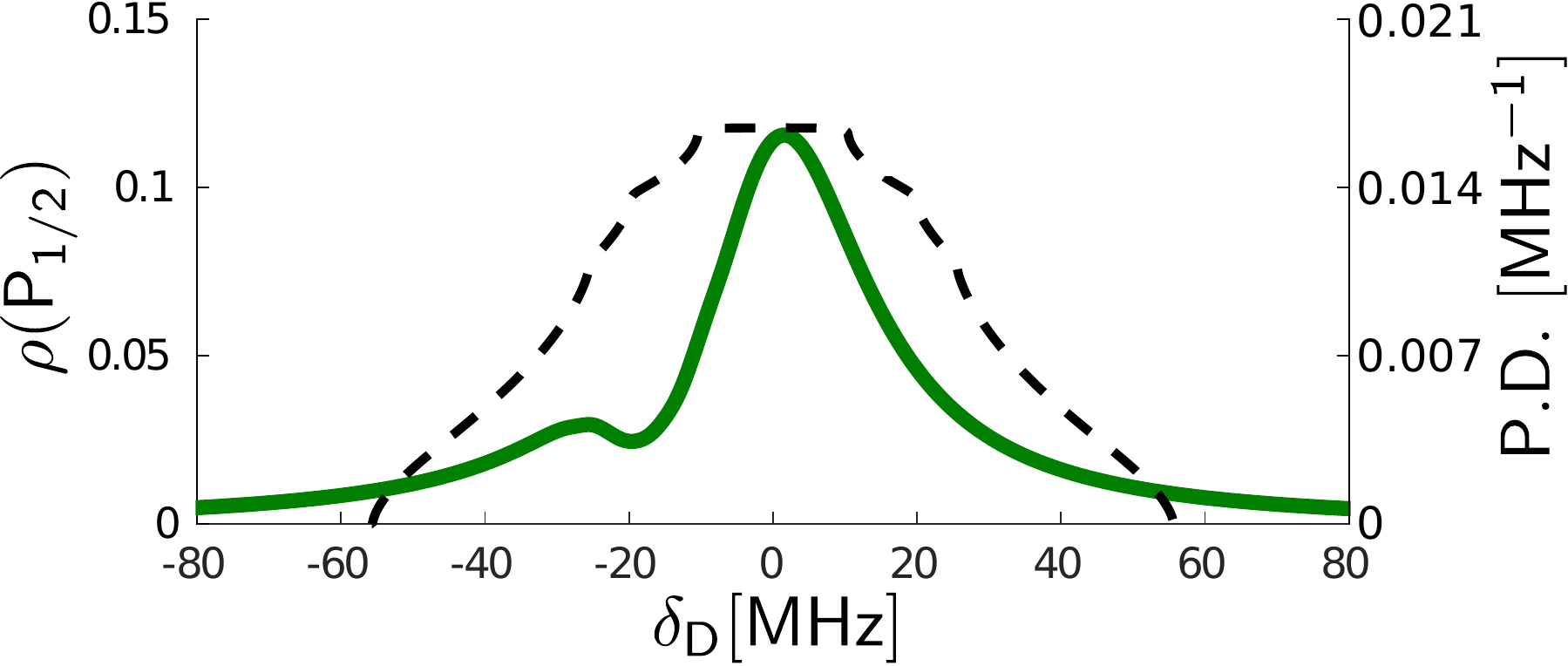}
\caption{\label{FIG. 22.}Excited state population $\rho(P_{1/2})$ as a function of the Doppler shift, $\delta_{D}=\delta_{422}=\frac{422}{1092}\delta_{1092}$. Laser parameters are same as in the \autoref{FIG. 2.}. The dashed line represents a combined Doppler shift distribution, $P_{3D}$, for an ion with 1 K energy in each axis ($E_{x}=E_{y}=E_{z}=1$ K$\cdot\textrm{k}_\textrm{B}$).}
\end{figure}

Since the total Doppler shift, $\delta_D=\delta^{x}+\delta^{y}+\delta^{z}$, for a 3D harmonic oscillator is a scalar quantity, it is convenient to write a single distribution of the Doppler shift in 3D. Following the derivation of \cite{Wesenberg2007} we convolve the $P_{1\textrm{D}}^i$ for all axis and obtain a 3D Doppler shift distribution $P_{3\textrm{D}}=P_{1\textrm{D}}^{x}\ast P_{1\textrm{D}}^{y}\ast P_{1\textrm{D}}^{z}$. The dashed line in \autoref{FIG. 22.} shows the convolved 3D Doppler shift distribution for an ion with a 1$\mathrm{\ K \cdot k_B}$ energy in each direction.

The scattering rate $\gamma$ of a hot ion is obtained by integrating the function $\gamma\left (\mathbf{par}, \delta_{D} \right )$ multiplied by $P_{3D}$ over the Doppler shift,
\begin{equation}
\gamma=\int_{-\delta_{max}}^{\delta_{max}} \gamma\left (\mathbf{par}, \delta_{D} \right )\cdot P_{3\textrm{D}}\left (\delta_{max}, \delta _{D} \right ) d\delta_{D}. \label{DNDT}
\end{equation}
Here, $\delta_{max}=\delta_M^x+\delta_M^y+\delta_M^z$ is determined by the energy of the ion. The average energy change per scattering event is $\hbar\mathbf{k_{422}}\cdot\mathbf{v}=\hbar\delta_{D}$. We use this fact and rewrite \autoref{delta2} as,
\begin{widetext}
\begin{equation}
\tfrac{\mathrm{d} E_{i}}{\mathrm{d} t} \left (\delta _{M}^{x,y,z} \right )=-\hbar\iiint \delta^{i}\cdot\gamma\left ( \mathbf{par},\delta_{D}=\delta^x+\delta^y+\delta^z \right )\prod_{i=x,y,z}P_{1\textrm{D}}\left ( \delta _{M}^{i}, \delta ^{i} \right ) d\delta^{x}d\delta^{y}d\delta^{z}+\tfrac{2}{3}\tfrac{\left (\hbar k \right )^2}{2m}\gamma. \label{DEDT}
\end{equation}
\end{widetext}

\autoref{DNDT} and \autoref{DEDT} together determine the time-dependent fluorescence rate of the ion during Doppler cooling. While \autoref{DEDT} determines the average change in energy per unit of time, \autoref{DNDT} is used to calculate the scattering rate as a function of energy.

\autoref{DEDT} takes into account only the momentum transfer due to the cooling light because the scattering rate contains information only on the 422 nm scattering. The repump light also imparts momentum on the ion, however, this effect is smaller by a factor of $\frac{\lvert\mathbf{k_{1092}}\rvert}{\lvert\mathbf{k_{422}}\rvert}\frac{\mathrm{\Gamma_{1092}}}{\mathrm{\Gamma_{422}}}\approx0.02$. We can include this effect by scaling the $\gamma\rightarrow\gamma\cdot\frac{\mathrm{\Gamma_{1092}}}{\mathrm{\Gamma_{422}}}$ and $\delta^i\rightarrow\delta^i\cdot\frac{\lvert\mathbf{k_{1092}}\rvert}{\lvert\mathbf{k_{422}}\rvert}$ in \autoref{DEDT} and summing both contributions.

\subsubsection{Validity of the weak binding approximation}
The timescale, $\tau$, in which the excited state reaches steady state depends greatly on the laser parameters. From a numerical solution of the eight-level Bloch equations, we observe that by saturating the $D_{3/2}\rightarrow P_{1/2}$ transition we can reduce $\tau$ and hence regain the weak-binding limit (inset of \autoref{FIG. 4.}). The timescale also depends on laser detuning and polarization. It is important to verify the validity of the weak binding approximation before using the Doppler cooling model.
\begin{figure}
\includegraphics{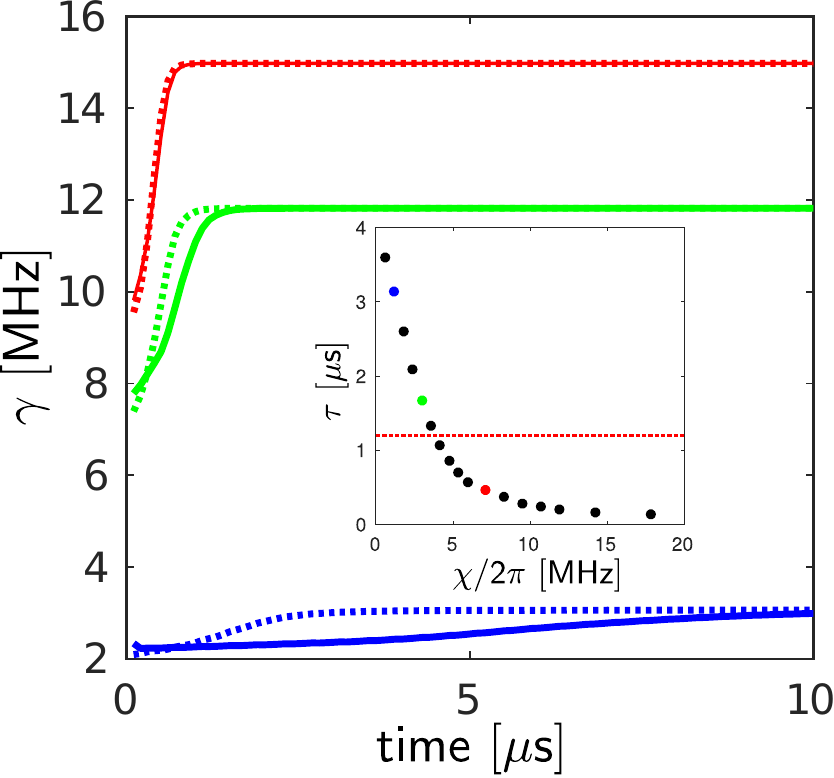}
\caption{\label{FIG. 4.}Comparison between the fluorescence rate predicted by OBS (solid lines) and Doppler cooling model (dashed lines). The initial energy in both methods is set to 1 K$\cdot$k$_\textrm{B}$ in the axial direction. We neglect recoil heating in this comparison. Different colors correspond to different repump Rabi frequencies. In the inset, we show the time scale in which the excited state reaches steady-state as a function of the repump Rabi frequency. Colored dots indicate the repump Rabi frequency used in the main figure. Red dashed line indicates the weak binding threshold: $\tau_{th}=\tfrac{2}{\omega}=1.2\mathrm{\mu{s}}$. Laser parameters are $\Omega_{422}/2\pi=12.2$ MHz, $\Omega_{1092}/2\pi=6.53$ MHz, $\delta_{422}=0$ MHz and $\delta_{1092}=-7.7$ MHz. The magnetic field, laser polarizations and linewidths are same as in \autoref{FIG. 2.}.}
\end{figure}

To validate our model, we compare it with the result of an optical Bloch simulation (OBS) in which we do not assume steady state \cite{Meir2017b}. Instead, we obtain the instantaneous scattering rate by propagating in time the 64 coupled dynamical Bloch equations of the 8-level system together with the equations of motion of the ion in the presence of a scattering force (see SM III.). The OBS is computationally very demanding and time-consuming which makes it impractical as a regular thermometry tool. Furthermore, the OBS does not provide any physical insight into the mechanisms that affect Doppler cooling. However, it does produce an exact result and can be used in regimes where the assumption of our model are no longer valid. 

\autoref{FIG. 4.} compares the photon scattering rate dynamics calculated by our model to the OBS result using different repump parameters. As expected, for weak repump intensities the weak-binding limit does not hold, and our model does not reproduce the OBS result, while for strong repump we see good agreement between the OBS and our model.

\subsubsection{Experimental verification of the model.}
To experimentally validate our model, we compare two different thermometry methods on a single $^{88}\mathrm{Sr}^{+}$ ion. We initialize the ion in a classical coherent state using an oscillating rf electric field drive, on-resonance with the axial trap frequency ($\omega_{ax}=417.5\ kHz$), with a constant rf drive power. We initialize the ion in different energies by changing the rf drive pulse length. 

\begin{figure*}
\includegraphics[width=\linewidth]{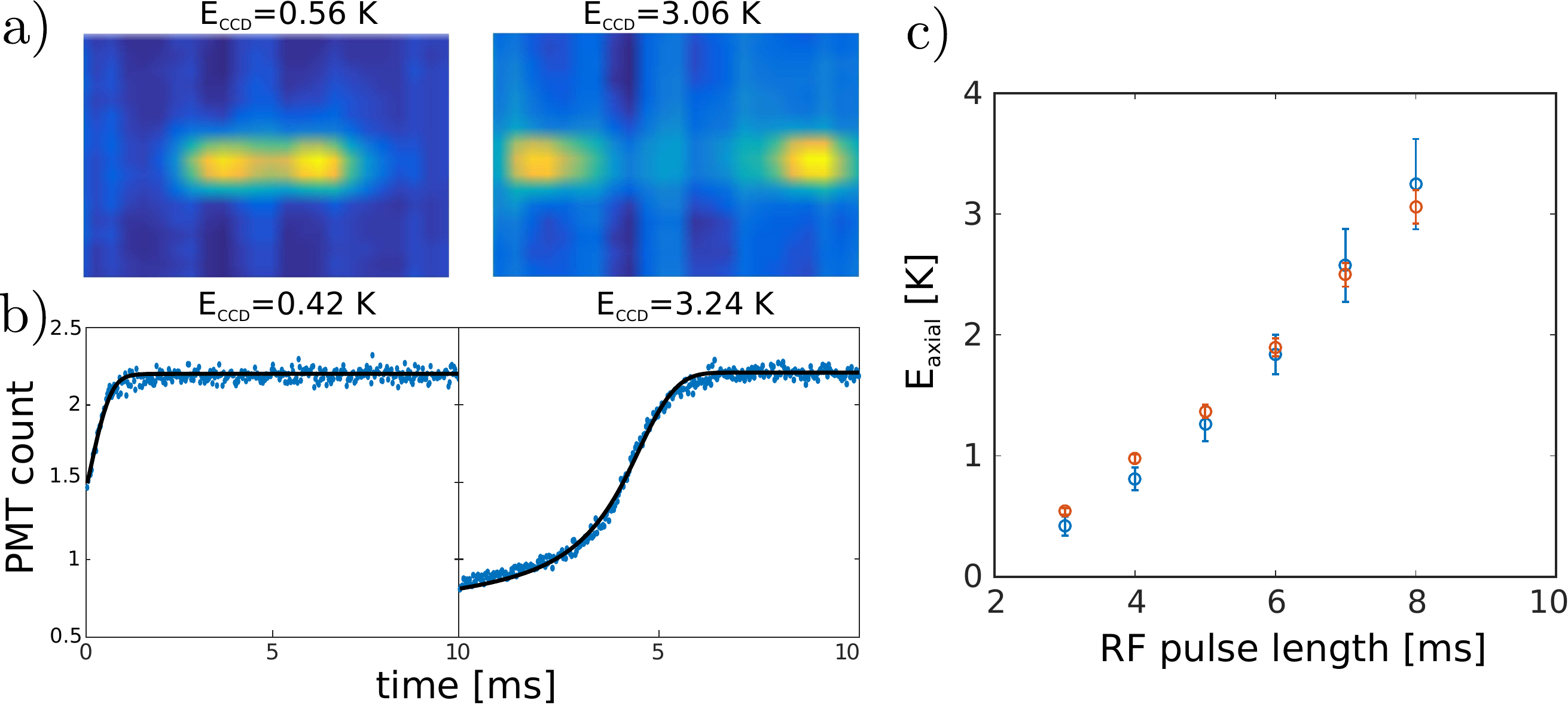}
\caption{\label{FIG. 5.}a) Time-averaged CCD images of an ion prepared in classical coherent states with different amplitudes. We extract the ions' energy by fitting the intensity profile to \autoref{PSP}. b) Experimentally measured ions' fluorescence during Doppler cooling (blue dots) for the same initialization as in a,b respectively. We extract the ion's energy from a single parameter fit (energy) to our model (black line). Laser parameters for the model are the same as in \autoref{FIG. 2.}. We set the 422 nm laser detuning to $\delta_{422}=-2.0$ MHz. c) A comparison between the two methods; Blue circles indicate energies derived from fitting the fluorescence signal of the Doppler cooling to our model and red circles are energies obtained from the intensity profile on the CCD. Error bars are 95\% confidence intervals.}
\end{figure*}

The first thermometry we use is the Doppler cooling thermometry. At the end of the ion's initialization, we turn on the cooling and repump lasers and monitor the ion's fluorescence for 10 ms with 30 $\mu$s binning. We repeat this process 200 times. In \autoref{FIG. 5.} we show the experimental results together with a single parameter fit (initial energy) to the Doppler cooling model. The good agreement between the model and the experimental time-dependent fluorescence signal suggests that the ion was prepared in a well-defined energy and that our model is valid.

We compare the energies that we obtained from the model with an alternative method for measuring the ions' energy when it is in a classical coherent state. In this method, we image the ion on a CCD and extract its energy from the shape of the intensity profile. For a 1D classical coherent state, the intensity profile is given by:
\begin{equation}
I\left(x\right)=\int_{0}^{\frac{2\pi}{\omega}} C\exp{\left(-\frac{\left(x-A\cos{(\omega t)}\right)^2}{2\sigma^2}\right)}\text{d}t. \title{PSP}
\end{equation}

Here, $\sigma$ is the width of our imaging system point-spread-function and $A$ is the ion's oscillation amplitude with frequency $\omega/2\pi$ from which we determine the energy of the ion: $E=\frac{1}{2}mA^2\omega^2$. We used on-resonance light such that Doppler shifts only reduce the scattering rate. In the analysis, we neglect the Doppler shift effect on the intensity profile since we are interested only in the oscillation amplitude. To prevent any mechanical effects of laser light on the energy of the ion, we image the ion with low laser intensity ($\Omega_{422}\approx0.1$ MHz) during 10 ms CCD exposures and repeat the measurement 5000 times to improve the signal-to-noise ratio. The experimental images are also shown in \autoref{FIG. 5.}.

A comparison between the energy extracted from the Doppler cooling fit and the CCD images for different energies are shown in \autoref{FIG. 5.}. The initial energies of the ion obtained by the two methods agree well. 
\onecolumngrid

\newpage
\twocolumngrid
\subsection{Paul trap}
In the preceding sections, we discussed the dynamics of Doppler cooling in a true harmonic potential while neglecting the oscillating fields of the ion's Paul trap. In linear Paul traps only one axis is static. In the remaining axes, the motion of the ion is a superposition of a slow frequency ($\omega/2\pi$) and fast frequency ($\Omega_{rf}/2\pi$) motion,
\begin{equation}
x_i(t)=\left [A_{i}^{emm}+A_{i}^{sec}\cos\left (\omega_i t \right ) \right ]\left [1+\tfrac{q_i}{2}\cos(\Omega_\textrm{rf} t) \right ].\title{MATHIEU}
\end{equation}
Here, $|q_i|<1$ is the Mathieu parameter which nulls for the static axis and has a finite value for the other two axes. $A_i^{sec}$ is the harmonic motion amplitude which is now superimposed with a fast modulation. We refer to this motion as inherent micromotion since it is proportional to the ions harmonic amplitude. The displacement of the ion from the rf null due to stray electric fields is $A_i^{emm}$. This displacement results in excess-micromotion with an amplitude $q_iA_i^{emm}/2$. While excess-micromotion is an artifact in Paul traps and can be reduced to negligible values \cite{Keller2015} the inherent part of micromotion is an intrinsic aspect of the ion's motion.

Because the micromotion velocity typically changes faster than the time in which the internal states of the atom reach steady-state, the absorption and emission of a photon can not be localized in phase space, and the weak binding approximation breaks. The analysis of the cooling process, therefore, requires a more involved approach.

\subsubsection{Micromotion sidebands}
Micromotion affects Doppler cooling by modifying the spectrum. To evaluate the effect of micromotion on the scattering rate we move to the micromotion reference frame. In this frame, the ion still undergoes a secular motion, but there is no micromotion. The electric field in the new reference frame is modulated at the rf frequency, $\Omega_\textrm{rf}$. 

In the case of excess micromotion, the ion senses additional sidebands with frequencies $\omega_{laser}\pm n\Omega_\textrm{rf}$, where $n$ is an integer number. We express the electric field in the presence of excess micromotion as,
\begin{multline}
\mathbf{E}\left ( t \right )=\mathbf{E_0} e^{i\left (\mathbf{k}\cdot\left[\mathbf{x_0}(t)+\tfrac{q}{2}\mathbf{A^{emm}}\cos\left(\Omega_\textrm{rf}t\right)\right]-\omega_{laser}t\right ) }=\\\mathbf{E_0} e^{\left [i\mathbf{k}\cdot\mathbf{x_0}(t)\right ]}\sum_{n}J_{n}\left(\beta^{emm}\right)e^{i \left(n\left(\Omega_\textrm{rf}t+\pi/2\right)-\omega_{laser}t\right).}
\label{elektrik}
\end{multline}
Here, $\mathbf{x_0}(t)$ is the secular and inherent part of the ion's motion and $\beta^{emm}=\frac{1}{2}\sum_{i=x,y}q_ik_iA_{i}^{emm}$ is the excess-micromotion modulation index. To first order in $\beta^{emm}$, excess-micromotion adds two sidebands to the ion spectrum at $\omega_{laser} \pm (\Omega_\textrm{rf})$. 

The inherent micromotion part of $\mathbf{x_0}(t)$ modulates the electric field twice, once in the harmonic trap frequency $\omega_i$ and second in the rf frequency $\Omega_\textrm{rf}$. The electric field in the presence of inherent micromotion can be written as:
\begin{multline}
\mathbf{E_0} e^{\left [i\mathbf{k}\cdot\mathbf{x_0}(t)\right ]}=\\
\mathbf{E_0} \prod_i e^{i k_i \left(A_i^{sec}\cos{\left(\omega_i t\right)} +A_i^{sec}\tfrac{q_i}{2}\cos{\left(\omega_i t\right)} \cos{\left(\Omega_\textrm{rf}t\right)}\right)}=\\
\mathbf{E_0} \prod_i e^{i k_i \left(A_i^{sec}\cos{\left(\omega_i t\right)} +A_i^{sec}\tfrac{q_i}{4}\left(\cos{\left(\left(\Omega_\textrm{rf}+\omega_i\right) t\right)} + \cos{\left(\left(\Omega_\textrm{rf}-\omega_i\right)t\right)}\right)\right)}.
\label{elektrik2}
\end{multline}
An expansion to Bessel functions can be performed here as well (see SM V.). To first order in the inherent micromotion modulation index, $\beta^{imm}_i=\frac{1}{4}q_ik_iA_{i}^{sec}$, this modulation adds four more additional sidebands at $\pm (\Omega_\textrm{rf} \pm \omega_i)$ for each of the radial modes, $i=x(y)$. The modified spectrum is obtained by recalculating the spectrum with amplitudes of sidebands which are obtained from the Bessel series expansion as we did in \autoref{elektrik}. The relative intensities of the inherent micromotion sidebands in first order can be expressed as a square of electric field components at frequency $\omega_{laser}\pm\Omega_\textrm{rf}\pm\omega_{x(y)}$ (for more details see SM V. \cite{Oberst1999}). The scattering rate is calculated by using the modified spectrum in \autoref{DNDT}. 

\subsubsection{Ion thermometry}
In the previous section, we have shown how micromotion affects the scattering rate by introducing motional sidebands. Here we will examine the effect of these sidebands on Doppler cooling fluorescence signal.

\begin{figure}
\includegraphics[width=\linewidth]{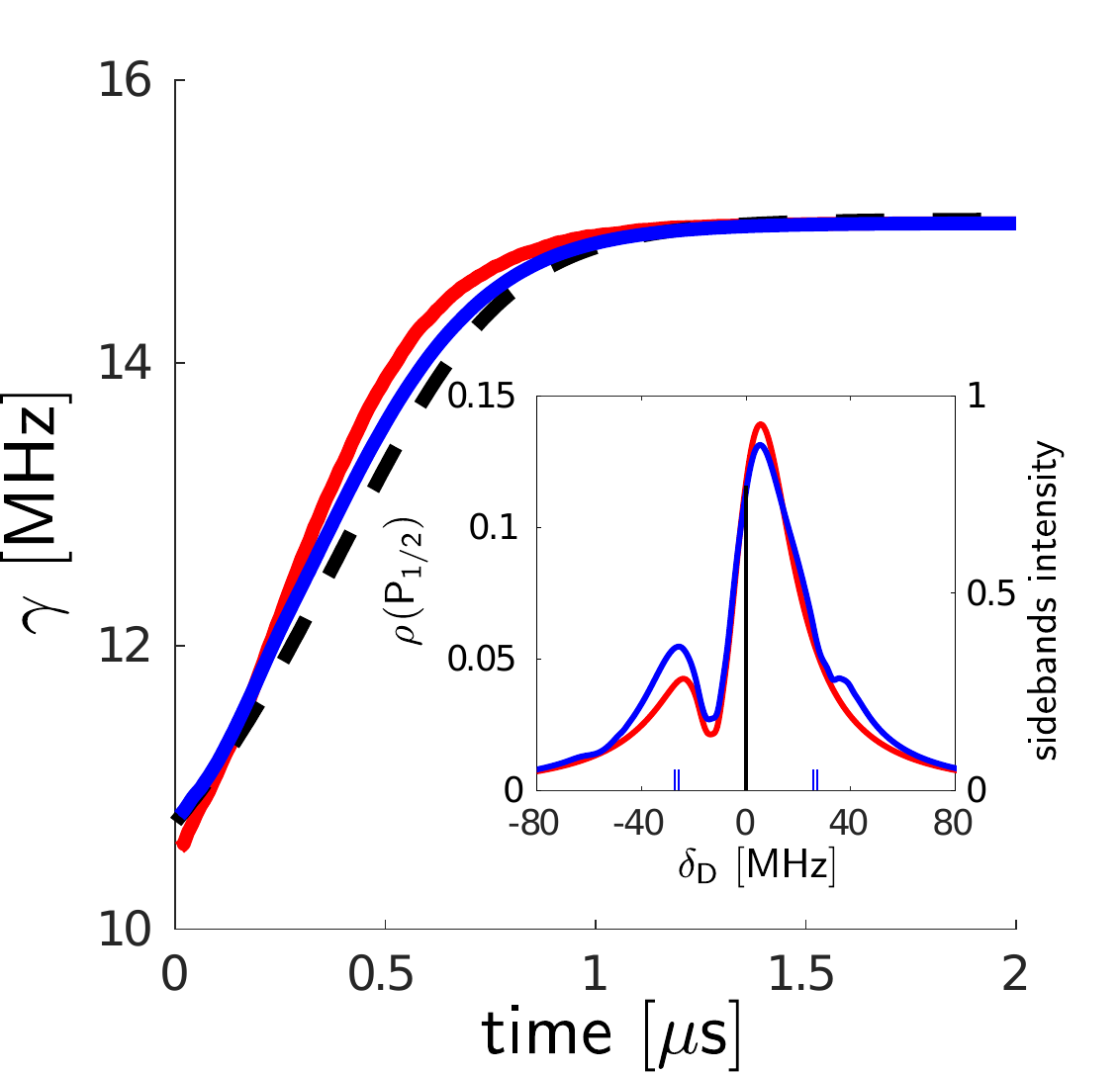}
\caption{\label{FIG. 6.}The effect of micromotion sidebands on the Doppler cooling fluorescence signal. Our model including $1^{st}$-order micromotion sidebands (blue) is in good agreement with the OBS result (dashed black curve). The red curve is calculated without micromotion sidebands. The inset shows the effect of these sidebands on the spectrum. The red curve is the ions' spectrum without sidebands. The blue curve is the spectrum including micromotion sidebands. Vertical lines represent the position of the $422\ \mathrm{nm}$ carrier (black) and inherent micromotion (blue) sidebands. The height of the lines is proportional to their relative power $J^2$. The ion's energy is set to $E_y=1\ \textrm{K}\cdot\textrm{k}_\textrm{B}$, $E_x=E_z=0.001\ \textrm{K}\cdot\textrm{k}_\textrm{B}$. Laser parameters are the same as in Section C.}
\end{figure}

Similarly, to what we did in section C.2, we compare our results to those of OBS. Extension of OBS from harmonic trap to Paul trap is straightforward (see SM III.). In \autoref{FIG. 6.} we compare the time evolution of fluorescence under Doppler cooling of an ion with initial energy $E=1\ \textrm{K}\cdot\textrm{k}_\textrm{B}$ in the $y$ direction calculated using OBS, and our model with and without including first-order micromotion sidebands. As seen, incorporating micromotion sidebands is necessary for our model to better match with the OBS result. With the inclusion of higher-order sidebands, agreement is expected to improve. The need to include higher order sidebands increases with higher initial energies.

\begin{figure}
\includegraphics[width=\linewidth]{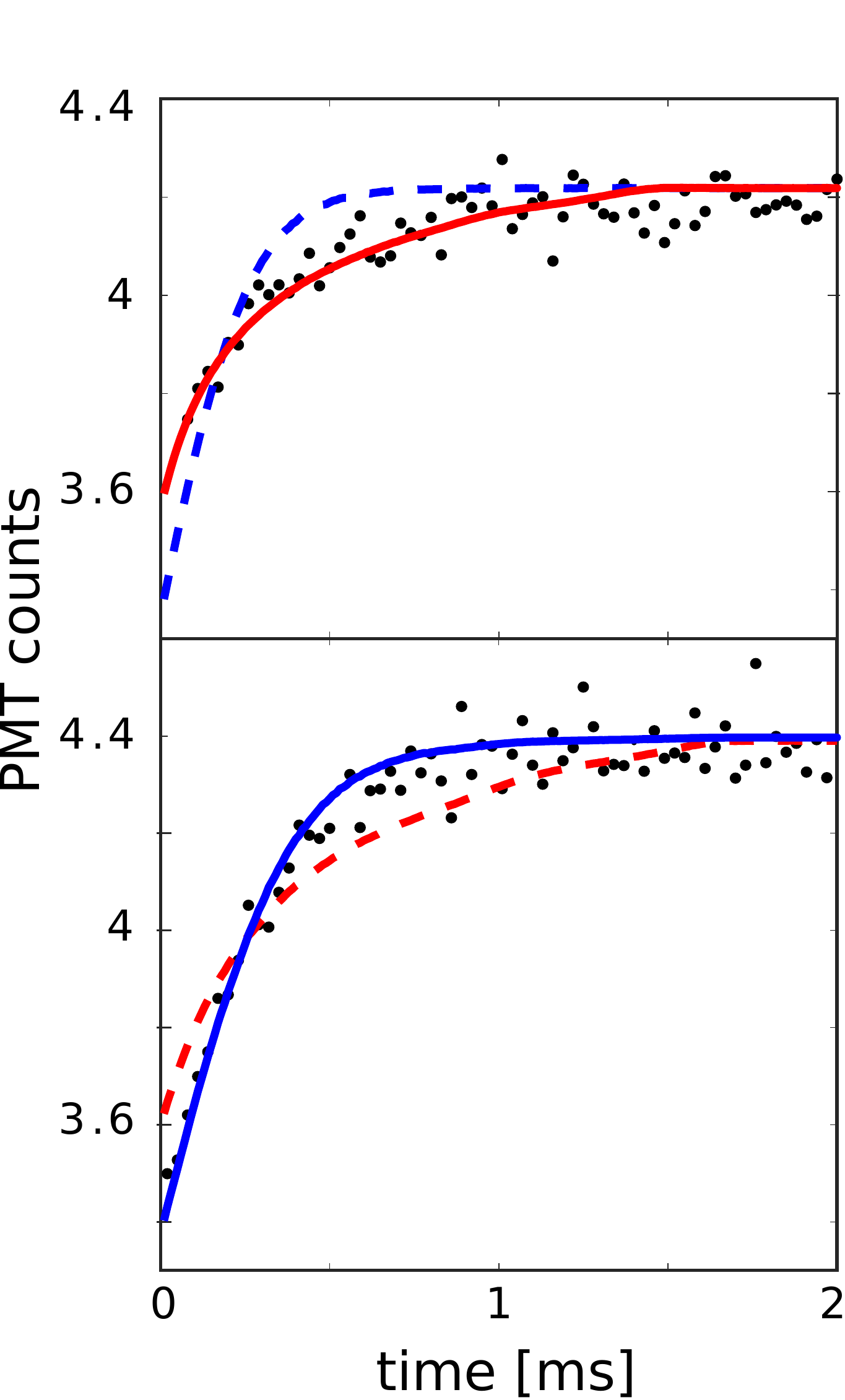}
\caption{\label{FIG. 7.}Fluorescence during Doppler cooling for different energy distributions together with a fit to a model. Data points (black) are number of photons collected at 50 $\mu$s intervals and averaged over 200 experimental realizations. Upper: Ion was prepared in a non-thermal Tsallis distribution. Lower: Ion was prepared in a thermal state. The curves are single parameter fits of the scattering rate assuming the Tsallis distribution (red) with $n=4$ and Maxwell-Boltzmann distribution (blue). Experimental parameters for this experiment can be found in Figure 3 of the SM. 422 nm laser detuning is $\delta_{422}=-18.57\pm 0.059$ MHz.}
\end{figure}
As an application of our model we determine the energy distribution of an ion in two different cases. Generally, the ion is not found in a specific energy state but rather in a statistical distribution of energies, $P\left(E\right)$. The averaged scattering rate is obtained from our model by weighting the fluorescence curves according to the distribution,
\begin{equation}
\left.\left\langle\frac{\text{d}N}{\text{d}t}\right\rangle=\int P\left(E\right)\frac{\text{d}N}{\text{d}t}\right\rvert_{E_0=E}.
\end{equation}

Here, we test our model with two different distributions (\autoref{FIG. 7.}). The Tsallis distribution, $P(E)\propto E^2/(1+E/n\text{k}_\text{B}T)^n$, which evolves after the ion collides multiple times with ultracold atoms \cite{DeVoe2009,Zipkes2011,Chen2014}. In our example, the energy scale, $\mathrm{k_B}\cdot T$, is determined by the intense excess micromotion, $E_{emm}=250\ \mathrm{mK\cdot k_B}$, induced in this experiment. The power-law, $n=4$, of the distribution is determined from a molecular-dynamics simulation \cite{Meir2016,Meir2017a}. The second distribution we study is thermal, $P(E)\propto E^2 e^{-E/\text{k}_\text{B}T}$. This distribution is produced by applying white voltage noise to one of the trap electrodes. We determine the heating rate due to this white noise using carrier Rabi spectroscopy on a narrow linewidth transition to be $305\ \mathrm{mK \cdot k_B/sec}$ (SM IV.).

We use our model to extract the distribution parameters from the experimental fluorescence curves. For the ion with a Tsallis energy distribution (red in \autoref{FIG. 7.}) the cooling rate is slower due to the broader energy distribution. We extract the ion's "temperature" \cite{Meir2016}, $T_\text{ion}=Tn/(n-2)$ from a fit to our model. The temperature, $T_\text{model}=155 \pm 13\ \mathrm{mK}$ agrees with the molecular dynamics simulation results, $T_\text{sim}=0.62*E_{emm}+7=162\ \mathrm{mK}$. The scaling is slightly different than in the referenced article due to different trap parameters used in this experiment \cite{Meir2016}. For the ion with a Maxwell-Boltzmann thermal distribution, the cooling rate is faster due to the relatively narrower energy distribution. We extract the ions' temperature $T_\text{model}=1080 \pm 50\ \mathrm{mK}$ using our model. This temperature agrees with the heating rate extrapolation of $T_\text{extrap}=915 \pm 45\ \mathrm{mK}$ using Rabi thermometry \cite{Leibfried2003}.
\section{Conclusion}
In this paper, we developed a model of Doppler cooling of a single multi-level ion trapped in a rf Paul trap. Our analysis includes the effects of both excess and inherent micromotion. Our model is a simple tool for understanding Doppler cooling dynamics.
This model can be used for ion thermometry in the range of 10's mK to 10's K. Below 10 mK Doppler shifts are too small to result in sufficient changes to the scattering rate. Above 10's K more and more micromotion sidebands have to be included which complicates the calculation significantly. With a good experimental signal-to-noise ratio, our model can also be used to distinguish between different energy distributions. We have bench-marked our method using coherent, thermal and non-thermal Tsallis energy distributions for energies between 0.5 to 3 $\mathrm{K \cdot k_B}$, and obtained a good agreement with alternative measurement methods and simulations. Doppler thermometry is a practical method because it requires only the same lasers used for Doppler cooling. It is an important tool for studying non-linear dynamics in ion crystals \cite{Lin2011},transport of ions \cite{Blakestad2009,Walther2012,Huber2008} or atom-ion collisions \cite{Meir2016,Zipkes2010a,Zipkes2010}. It can also serve as an important method for future thermodynamics experiments in ion traps \cite{Roßnagel2016,Abah2012}.

This work was supported by the Crown Photonics
Center, ICore-Israeli excellence center circle of light, the
Israeli Science Foundation, the U.S.-Israel Binational
Science Foundation, and the European Research Council (consolidator grant 616919-Ionology). We thank Asaf Miron for helpful conversations and for remarks on the manuscript.

\widetext
\clearpage
\begin{center}
	\textbf{\Large Supplementary material: Doppler cooling thermometry of a multi-level ion in the presence of micromotion}
	\end{center}
\setcounter{equation}{0}
\setcounter{figure}{0}
\setcounter{page}{1}
\setcounter{section}{0}
\setcounter{tocdepth}{0}
\smallskip
\begin{center}
	\textbf{\large I. Experimental apparatus}
\end{center}
\smallskip
We perform our experiments in a linear segmented rf ($\Omega_\textrm{rf}=26.51 \ \mathrm{MHz}$) Paul trap with secular frequencies $\omega_{x,y,z}=[0.42\ 0.73\ 0.99]\ \mathrm{MHz}$. The magnetic field in the center of the trap is 3$\pm$0.02 Gauss. We derive the magnetic field from a spectroscopic scan of the different Zeeman transitions between the S$_{1/2}\rightarrow$D$_{5/2}$ levels (Figure 2 in the main text). Beams polarizations are 0 and 45 degrees with respect to the magnetic field for the 422 nm and 1092 nm lasers respectively. Both beams k-vectors are perpendicular to the magnetic field. We measure the collection efficiency (Photons collected/Photons scattered) of violet photons of the PMT to be $1/201\pm5$, by shelving to the D$_{5/2}$ level and then applying a repump pulse. This way the ion is emitting a single photon each time and we can measure the detection probability. We measured the angles between the lasers $k$-vectors and the trap axes by comparing carrier and sideband Rabi frequencies when the ion is cooled near the ground-state, $k_{x,y,z}=|k|.[\cos{(46)},\cos{(62.5)},\cos{(56.5)}]$. For the experiment of figure 7 in the main text, the lasers beams polarizations were 6 and 35 degrees to the magnetic field and the collection efficiency was $1/190\pm5$.
\bigskip
\begin{center}
	\textbf{\large II. The interaction of an eight level system with two coherent light fields}
\end{center}
\medskip
\begin{figure}[H]
	\centering
	\includegraphics{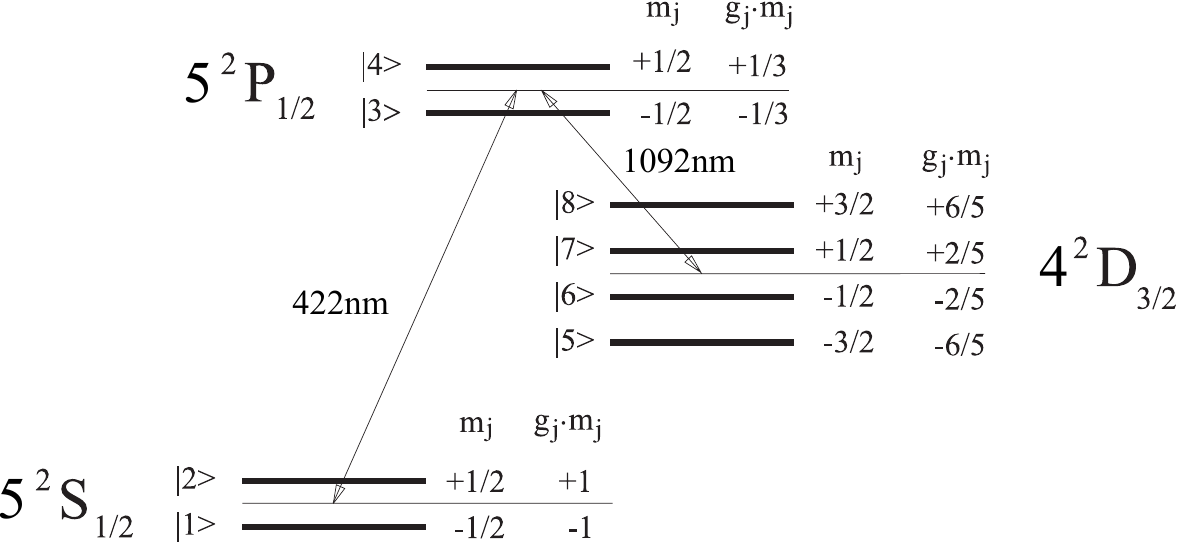}
	\caption{\label{FIG. 1.}The level scheme of $^{88}$Sr$^+$ with the Zeeman splitting relevant to the Doppler cooling model. $g_j$ is the Lande factor.}
\end{figure}
In this section and also in section V we follow the derivation of \cite{Oberst1999}. The Lindblad master equation gives the time evolution of the density operator $\hat{\rho}$:
\begin{equation}
\begin{split}
&\frac{\text{d}\hat{\rho}}{\text{d}t}=\mathcal{L}{\left[\hat\rho\right]},\\
&\mathcal{L}=-\frac{i}{\hbar}\left[\hat{H},\hat{\rho}\right]+\hat{D}. \label{DEDT}
\end{split}
\end{equation}
Here, $\hat{H}$ is the Hamiltonian of the system and the $\hat{D}$ is a Lindblad operator describing non-unitary processes. In our case these are spontaneous decay and finite laser linewidths. Decay occurs from the P level to the S level and from the P level to the D level. The dissipative operator $\hat{D}$ has the form:
\begin{equation}
\hat{D}=-\frac{1}{2}\sum_{i}\hat{C}_i^{\dagger}\hat{C}_i\rho+\rho\hat{C}_i^{\dagger}\hat{C}_i-2\hat{C}_i^{\dagger}\rho\hat{C}_i.
\end{equation}
The damping terms, $\hat{C}_{1:6}$, describe spontaneous decay from P level:
\begin{equation}
\begin{aligned}[c]
&\hat{C}_1=\sqrt{\frac{2}{3}\Gamma_{P\rightarrow{S}}}|1\rangle\langle4|,\\ &\hat{C}_2=\sqrt{\frac{2}{3}\Gamma_{P\rightarrow{S}}}|2\rangle\langle3|,\\ &\hat{C}_3=\sqrt{\frac{\Gamma_{P\rightarrow{S}}}{3}}\left(|1\rangle\langle3|-|2\rangle\langle4|\right),\\
\end{aligned}
\begin{aligned}[c]
&\hat{C}_4=\sqrt{\frac{\Gamma_{P\rightarrow{D}}}{6}}\left(\sqrt{3}|5\rangle\langle3|+|6\rangle\langle4|\right),\\ &\hat{C}_5=\sqrt{\frac{\Gamma_{P\rightarrow{D}}}{6}}\left(|7\rangle\langle3|+\sqrt{3}|8\rangle\langle4|\right),\\ &\hat{C}_6=\sqrt{\frac{\Gamma_{P\rightarrow{D}}}{3}}\left(|6\rangle\langle3|+|7\rangle\langle4|\right).
\end{aligned}
\end{equation}
Here, $|i\rangle$ are the eigenstates of energy levels in the Fock state basis (\autoref{FIG. 1.}).
$\Gamma_{P\rightarrow{S}}=128$ MHz and $\Gamma_{P\rightarrow{D}}=7.46$ MHz \cite{Zhang2016}.
The effect of finite lasers linewidths is described by:
\begin{equation}
\begin{split}
&\hat{C}_7=\sqrt{2\Gamma_{422}}\left(|1\rangle\langle1|+|2\rangle\langle2|\right),\\ &\hat{C}_8=\sqrt{2\Gamma_{1092}}\left(|5\rangle\langle5|+|6\rangle\langle6|+|7\rangle\langle7|+|8\rangle\langle8|\right).
\end{split}
\end{equation}
Here, $\Gamma_{422}$ and $\Gamma_{1092}$ express the cooling and repump laser linewidths which are on the order of 100's of kHz.

The atomic part of the Hamiltonian has the form:
\begin{equation}
\hat{H}_{atom}=\sum_i^8\hbar\omega_i|i\rangle\langle i|=\sum_{i=1}^2\hbar\omega_S|i\rangle\langle i|+\sum_{i=3}^4\hbar\omega_P|i\rangle\langle i|+\sum_{i=5}^8\hbar\omega_D|i\rangle\langle i|.
\end{equation}
For co-propagating laser beams that are perpendicular to the magnetic field, the part of the Hamiltonian, $\hat{H}$, that describes the coupling between all eight eigenstates by two laser lights, have the following matrix elements:
\begin{equation}
\begin{aligned}[c]
&\hat{H}_{1,3}=+\frac{1}{\sqrt{3}}\Omega_{422}\cos{(\alpha)}e^{i\omega_{422}t},\\
&\hat{H}_{2,3}=-\frac{1}{\sqrt{3}}\Omega_{422}\sin{(\alpha)}e^{i\omega_{422}t},\\
&\hat{H}_{5,3}=-\frac{1}{2}\Omega_{1092}\sin{(\beta)}e^{i\omega_{1092}t},\\
&\hat{H}_{6,3}=-\frac{1}{\sqrt{3}}\Omega_{1092}\cos{(\beta)}e^{i\omega_{1092}t},\\
&\hat{H}_{7,3}=+\frac{1}{2\sqrt{3}}\Omega_{1092}\sin{(\beta)}e^{i\omega_{1092}t},\\
&\hat{H}_{8,3}=0,
\end{aligned}
\begin{aligned}[c]
&\hat{H}_{1,4}=-\frac{1}{\sqrt{3}}\Omega_{422}\sin{(\alpha)}e^{i\omega_{422}t},\\
&\hat{H}_{2,4}=+\frac{1}{\sqrt{3}}\Omega_{422}\cos{(\alpha)}e^{i\omega_{422}t}\\
&\hat{H}_{5,4}=0,\\
&\hat{H}_{6,4}=-\frac{1}{2\sqrt{3}}\Omega_{1092}\sin{(\beta)}e^{i\omega_{1092}t},\\
&\hat{H}_{7,4}=-\frac{1}{\sqrt{3}}\Omega_{1092}\cos{(\beta)}e^{i\omega_{1092}t},\\
&\hat{H}_{8,4}=+\frac{1}{2}\Omega_{1092}\sin{(\beta)}e^{i\omega_{1092}t}.
\end{aligned}
\end{equation}
Here, the Rabi frequencies, $\Omega_{1092}$ and $\Omega_{422}$ are defined as $\Omega=\frac{E\cdot D}{2\hbar}$. $\alpha$ and $\beta$ are the linear polarization angles of the 422 nm and 1092 nm beams, respectively, to the magnetic field axis. $\omega_{422}$ and $\omega_{1092}$ are the laser beams frequencies. 
Finally, we move to the interaction representation using the unitary operator:
\begin{equation}
\hat{U}=\sum_{i=1}^2e^{-i\omega_{422}t}|i\rangle\langle{i}|+\sum_{i=5}^8e^{-i\omega_{1092}t}|i\rangle\langle{i}|.
\end{equation}
In the interaction representation, $\hat{H}$ and $\rho$ transform according to:
\begin{equation}
\hat{H}'=\hat{U}\hat{H}\hat{U}^{\dagger}-i\hbar\hat{U}\frac{\text{d}\hat{U}^{\dagger}}{\text{d}t}, \label{H}
\end{equation}
\begin{equation}
\rho'=\hat{U}\rho\hat{U}^{\dagger}.
\end{equation}
The detunings are now included in $\hat{H}'$,
\begin{equation}
\begin{aligned}
&\Delta_{422}=\omega_{422}-(\omega_{P}-\omega_{S}),\\
&\Delta_{1092}=\omega_{1092}-(\omega_{P}-\omega_{D}).
\end{aligned}
\end{equation}
To obtain the optical Bloch equations, we rewrite $\hat{\rho}'$ into a vector form: 
\begin{equation}
\rho'=(\rho_{11},\rho_{12},...,\rho_{87},\rho_{88})
\end{equation}
In the vector notation we omitted the prime for clarity. The time evolution of the density matrix is now given by:
\begin{equation}
\frac{\text{d}\rho_{rs}}{\text{d}t}=\sum_{kj}L_{rs,kj}\rho_{kj}.\label{time}
\end{equation}
Here, L is the Liouville matrix that is given by:
\begin{equation}
L_{rs,kj}=-\frac{i}{\hbar}\left(\mathcal{H}_{rk}\delta_{js}-\mathcal{H}_{js}^{\dagger}\delta_{rk}\right)+\sum_m\left(C_m\right)_{rk}\left(C_m^{\dagger}\right)_{js},
\end{equation}
where $\mathcal{H}=\hat{H'}-\frac{i}{2\hbar}\sum_m\hat{C}_m^{\dagger}\hat{C}_m$.

To obtain the spectrum, the steady-state solution is required. This is obtained by solving the equation: $\sum_{kj}L_{rs,kj}\rho_{kj}=0$.
\bigskip
\begin{center}
	\textbf{\large III. Cooling Dynamics Simulation}
\end{center}
\medskip
We propagate the ion motion in position and velocity coordinates:
\begin{equation}
\frac{\text{d}x_i}{\text{d}t}=v_i
\end{equation}
\begin{equation}
\frac{\text{d}v_{i}}{\text{d}t}=-x_{i}(a_i+2q_i\cos(\Omega_\textrm{rf}t))\Omega_\textrm{rf}^2/4+\hbar\large \rho_{\small P}\normalsize\left(\Gamma_{P\rightarrow S}k_{i,422}+\Gamma_{P\rightarrow D}k_{i,1092}\right)/m.
\end{equation}
Here, $a_i$ and $q_i$ are the Paul trap Mathieu parameters, $\Omega_\textrm{rf}$ is the trap rf frequency, $k_{i,422}$ and $k_{i,1092}$ are the projection of the 422 and 1092 lasers k-vector on the i-th mode respectively, $m$ is the ion mass and $\rho_{\small P}=\rho_{33}+\rho_{44}$ is the excited state population. The damping term takes into account the mechanical action of the lasers on the ion motion.
Simultaneously, we also propagate the ion's density matrix using \autoref{time}.

\begin{equation}
\frac{\text{d}\rho_{rs}}{\text{d}t}=\sum_{kj}L_{rs,kj}\rho_{kj}.
\end{equation}
The ion motion and the density matrix are coupled since the Liouville operator depends on the ion's velocity through the lasers detuning and the ion's velocity depends on the density matrix through the damping term. Such a simulation is computationally demanding and does not provide any physical insight into the mechanism of Doppler cooling. However, it produces an exact results \cite{Meir2017b}.
\bigskip
\begin{center}
	\textbf{\large IV. Rabi oscillations thermometry of a heated ion}
\end{center}
\medskip
\begin{figure}
	\centering
	\includegraphics{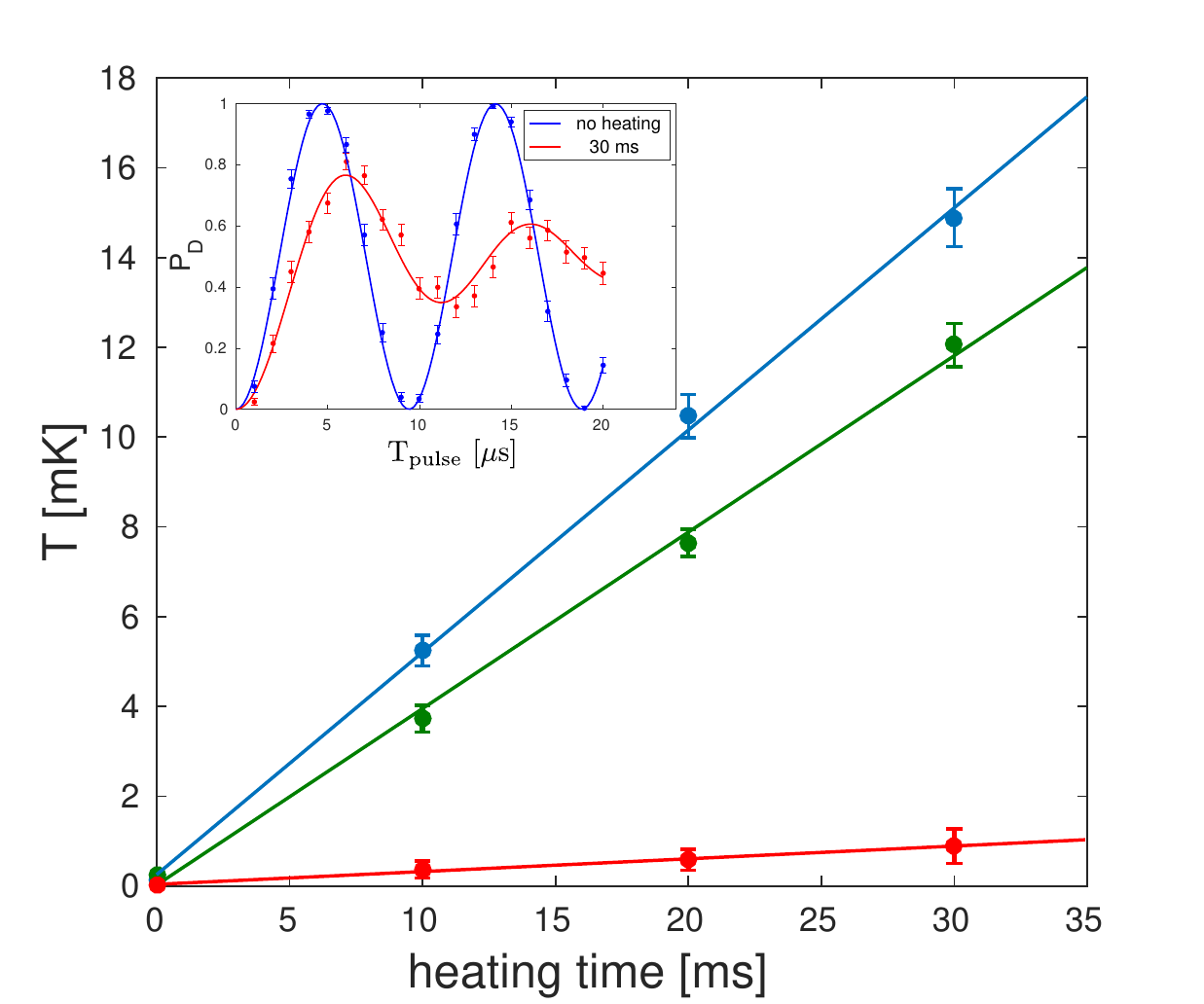}
	\caption{\label{FIG. 2.}Ion temperature of each mode measured as a function of the white-noise pulse duration. The solid lines represent a linear fit for each axis: x (red), y (green) and z (blue). The inset shows examples of carrier Rabi spectroscopy for a ground-state cooled ion (blue) and for z-axis after 30ms heating pulse. We scanned the shelving pulse time and measured the shelving probability $P_D$. Each data point corresponds to 200 repetitions. Error bars are binomial distribution standard deviation. We fitted the data assuming a thermal distribution in a single mode since other modes are kept near the ground-state.}
\end{figure}
In the main text we used white electric noise on the trap electrodes to heat the ion to several Kelvin. We calibrated the white-noise heating rate for each of the ion's modes using carrier Rabi spectroscopy (see \autoref{FIG. 2.}). We performed this spectroscopy at much shorter pulse times during which the ion heated up to temperatures up to 10 mK. We then linearly extrapolated the ion temperature for extended pulse times and compared this result to a direct measurement using Doppler cooling thermometry. The heating rate was measured by first preparing the ion in the ground state of all of its modes ($\bar{n}<0.1$). During heating, we stopped ground state cooling on the mode of interest but continued to ground state cool on the remaining two modes. The extracted heating rates are: $[28\pm 2, 393\pm 19, 495\pm 24]$ $\mathrm{mK.k_B/sec}$ for x, y and z axis respectively.
\bigskip
\begin{center}
	\textbf{\large V. Inclusion of micromotion Sidebands in the spectrum}
\end{center}
\medskip
To calculate the spectrum of an ion in the presence of a micromotion sidebands, we need to modify the detunings that appear after transforming the Hamiltonian into a rotating frame in \autoref{H}.
\begin{equation}
\begin{split}
&\delta_{422}=\omega_{422}-(\omega_P-\omega_S)\\
&\delta_{1092}=\omega_{1092}-(\omega_P-\omega_D)\\
\end{split}
\end{equation}
Sidebands appear because the laser detunings are modulated:
\begin{equation}
\begin{split}
&\delta_{422}=\delta_{422}^0+k_{422}v_x^{inh}\left\{\cos{(\left(\Omega+\omega_x \right)t)}+\cos{(\left(\Omega-\omega_x\right)t)}\right\}\\
&\delta_{1092}=\delta_{1092}^0+k_{1092}v_x^{inh}\left\{\cos{(\left(\Omega+\omega_x \right)t)}+\cos{(\left(\Omega-\omega_x\right)t)}\right\}\\
\end{split}
\end{equation}
Here we treat the case of inherent micromotion due to motion along the $x$ axis. $v_x^{inh}=\frac{q\Omega A_{x}^{sec}}{4}$

\begin{figure}
	\centering
	\includegraphics[scale=0.8]{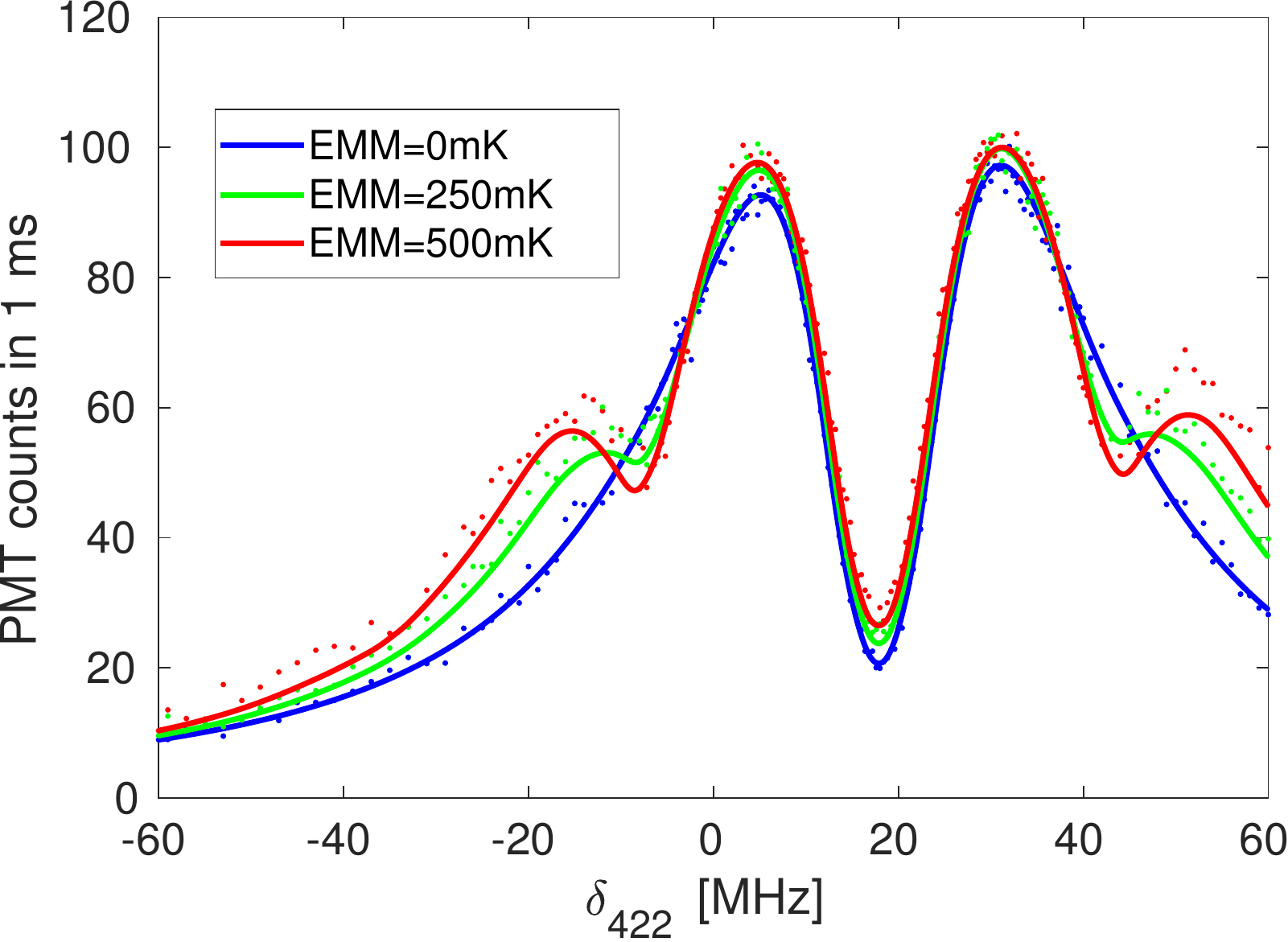}
	\caption{\label{FIG. 3.}A spectroscopic scan of the $S_{1/2}\rightarrow P_{1/2}$ transition for various EMM. Points are the number of photons detected on a Photon-counter which is proportional to the excited state population. The solid lines are fit to the steady-state solution of the 8-level optical Bloch equations. Experimental parameters are: $\Omega_{422}/2\pi=15.6\pm 0.05$ MHz, $\Omega_{1092}/2\pi=10.7\pm 0.22$ MHz, $\delta_{1092}=-0.66\pm 0.06$ MHz. These parameters are extracted from $EMM=0$ (blue) fit. Beams polarizations are 6 and 35 degrees to the magnetic field for the 422 nm and 1092 nm laser respectively.}
\end{figure}

Because $L$ is a linear matrix of the detunings, it is possible to separate the detuning by introducing unit Liouville matrices.
\begin{equation}
\begin{split}
&\Delta L_{422}=L\left(1,0,par\right)-L\left(0,0,par\right)\\
&\Delta L_{1092}=L\left(0,1,par\right)-L\left(0,0,par\right)\\
\end{split}
\end{equation}
Where $L\left(\delta_{422},\delta_{1092},par\right)$ is a function of both detunings and all other laser parameters which are fixed.
The modified Liouville matrix now gives the time evolution:
\begin{equation}
L=L_0+2\Delta L\left\{\cos{(\Omega+\omega_x)t}+\cos{(\Omega-\omega_x)t}\right\}
\end{equation}
\begin{equation}
\Delta L=\frac{v_x^{inh}}{2}\left\{k_{422}\Delta L_{1092}+k_{1092}\Delta L_{422}\right\}
\end{equation}
We will treat only first-order sidebands. In the long time we expect the solution with frequency components only at multiples of $\Omega+\omega_x$ and $\Omega-\omega_x$:
\begin{equation}
\rho=\rho_{0,0}+\rho_{1,0}e^{i\left(\Omega+\omega\right) t}+\rho_{-1,0}e^{-i\left(\Omega+\omega\right) t}+\rho_{0,1}e^{i\left(\Omega-\omega\right) t}+\rho_{0,-1}e^{-i\left(\Omega-\omega\right) t}
\end{equation}
Combining eq. (20, 18 and 11) we get following equations for $\rho_{i,i}$.
\begin{equation}
L_0\rho_{0,0}+\Delta L\left(\rho_{1,0}+\rho_{-1,0}+\rho_{0,1}+\rho_{0,-1}\right)=0
\label{rho1}
\end{equation}
\begin{equation}
\begin{split}
&\left(L_0-i\left(\Omega+\omega\right)\right)\rho_{1,0}+\Delta L\rho_{0,0}=0\\
&\left(L_0+i\left(\Omega+\omega\right)\right)\rho_{-1,0}+\Delta L\rho_{0,0}=0\\
&\left(L_0-i\left(\Omega-\omega\right)\right)\rho_{0,1}+\Delta L\rho_{0,0}=0\\
&\left(L_0+i\left(\Omega-\omega\right)\right)\rho_{0,-1}+\Delta L\rho_{0,0}=0\\
\end{split}
\end{equation}
We can now express $\rho_{i,j}$ in terms of $\rho_{0,0}$
\begin{equation}
\begin{split}
&\rho_{1,0}=-\Delta L\rho_{0,0}\left(L_0-i\left(\Omega+\omega\right)\right)^{-1}\\
&\rho_{-1,0}=-\Delta L\rho_{0,0}\left(L_0+i\left(\Omega+\omega\right)\right)^{-1}\\
&\rho_{0,1}=-\Delta L\rho_{0,0}\left(L_0-i\left(\Omega-\omega\right)\right)^{-1}\\
&\rho_{0,-1}=-\Delta L\rho_{0,0}\left(L_0+i\left(\Omega-\omega\right)\right)^{-1}\\
\label{rho2}
\end{split}
\end{equation}
Plugging \autoref{rho2} into \autoref{rho1} we can write the Liouville matrix as
\begin{equation}
L=L_0-\Delta L^2\left(\left(L_0-i\left(\Omega+\omega\right)\right)^{-1}+\left(L_0+i\left(\Omega+\omega\right)\right)^{-1}+\left(L_0-i\left(\Omega-\omega\right)\right)^{-1}+\left(L_0+i\left(\Omega-\omega\right)\right)^{-1}\right)
\end{equation}
To compare the spectra, we calculate including sidebands with experiment we fit our model to a spectrum without excess micromotion. Then, we recalculate the spectrum adding the excess micromotion and compare it to experimental spectra with excess micromotion \autoref{FIG. 3.} (EMM amplitude was determined using 674nm narrow transition sideband spectroscopy). As seen, there is good agreement between the spectra obtained using our model and those obtained experimentally.

In the treatment above we approximated the effect of the micromotion by including the first-order sidebands only. We now examine the validity of this approximation at different ion temperatures and EMM levels. The optical power in each sideband is proportional to the square of the electric field amplitude at its frequency. This amplitude can be obtained by expanding Equation 9 from the main text into a Bessel series.
$I_{i}^{j,k}$ is the electric field intensity at frequency of a corresponding $i$-th axis $|j|+|k|$ order sideband. The frequency can be expressed as $j\left(\Omega+\omega_i\right)+k\left(\Omega-\omega_i\right)$.
\begin{equation}
\begin{split}
&I_{x}^{\pm1,0}=\left(J_{\pm1}(\beta^{x})\cdot J_0(\beta^{x})\cdot J_0(\beta^{y})\cdot J_0(\beta^{y})\cdot J_0(\beta^{emm})\right)^2\\
&I_{x}^{0,\pm1}=\left(J_0(\beta^{x})\cdot J_{\pm1}(\beta^{x})\cdot J_0(\beta^{y})\cdot J_0(\beta^{y})\cdot J_0(\beta^{emm})\right)^2\\
&I_{y}^{\pm1,0}=\left(J_{\pm1}(\beta^{x})\cdot J_0(\beta^{x})\cdot J_{\pm1}(\beta^{y})\cdot J_0(\beta^{y})\cdot J_0(\beta^{emm})\right)^2\\
&I_{y}^{0,\pm1}=\left(J_0(\beta^{x})\cdot J_0(\beta^{x})\cdot J_0(\beta^{y})\cdot J_{\pm1}(\beta^{y})\cdot J_0(\beta^{emm})\right)^2
\end{split}
\end{equation}.

The carrier power is given by,
\begin{equation}
I^{0,0}=\left(J_0(\beta^{x})\cdot J_0(\beta^{x})\cdot J_0(\beta^{y})\cdot J_0(\beta^{y})\cdot J_0(\beta^{emm})\right)^2
\end{equation}.

Since the Doppler cooling thermometry is usually performed below saturation, it is important to verify whether calculating the spectrum up to 1st order micromotion sidebands is sufficient. As a figure of merit, we verify that the sum of the Bessel functions \autoref{bessel} up to first order is close to one. The sum of optical power vs. temperature is shown in \autoref{FIG. 4.}. As seen, up to a temperature of 1 K almost all optical power ($>$90\%) is included by the first sideband approximation. Above 1 K more sidebands have to be included. 
\begin{equation}\label{bessel}
I=I^{0,0}+2\cdot\left(I_{x}^{1,0}+I_{x}^{0,1}+I_{y}^{1,0}+I_{y}^{0,1}\right).
\end{equation}
\begin{figure}
	\centering
	\includegraphics[scale=0.9]{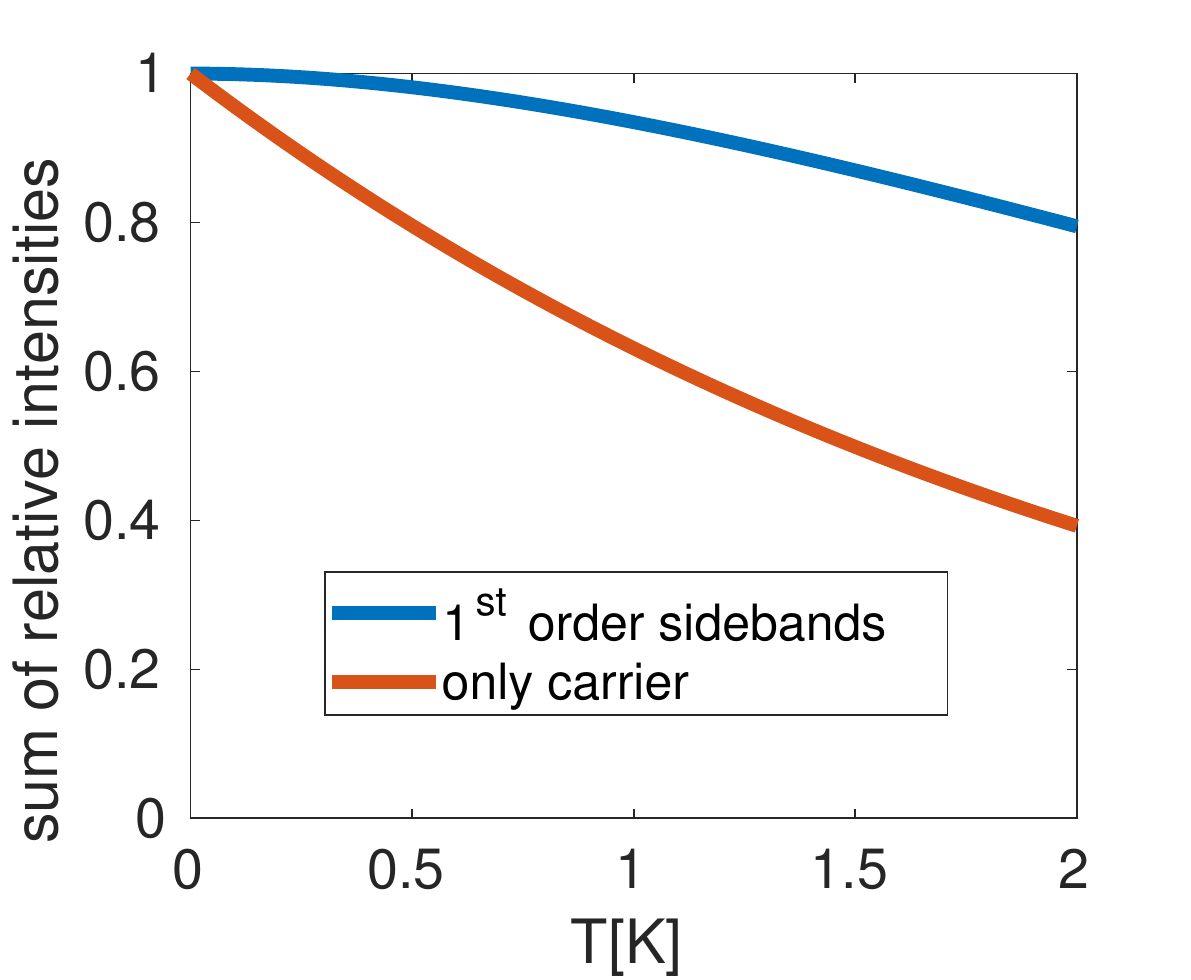}
	\caption{\label{FIG. 4.}Proportion of a laser field captured by zero order (red) and first order (red) Bessel series. We see that zero order becomes insufficient at temperatures above 0.2K while first order treatment captures most of the power ($>90\%$) at temperatures of 1K. Above 1K, second order sidebands are required. The trap frequencies are $\mathrm{\omega_x=0.73\ and\ \omega_y=0.99\ MHz}$.}
\end{figure}
\end{document}